\documentclass[aps,pra,twocolumn,groupedaddress]{revtex4-2}
\usepackage{eurosym}
\usepackage{amsmath}
\usepackage{mathrsfs}
\usepackage{graphicx}
\usepackage[normalem]{ulem}
\usepackage{color}
\usepackage{hyperref}

\begin{document}

\title{Phase transitions and dynamics of one-dimensional solitons in
spin-orbit-coupled Bose-Bose mixtures}

\author{Gui-hua Chen$^1$, Hongcheng Wang$^1$, Boris A. Malomed$^{2,3}$}
\author{Haiming Deng$^{4}$}
\thanks{~Corresponding author: \href{mailto:woshidenghaiming@126.com}%
	{woshidenghaiming@126.com}}
\author{Yongyao Li$^{5}$}
\affiliation{$^1$ Department of Electronic Engineering, Dongguan University 
	of Technology, Dongguan 523808, China}
\affiliation{$^2$ Tel Aviv University, Tel Aviv 69978, Israel}
\affiliation{$^3$ Instituto de Alta Investigaci\'{o}n, Universidad de 
	Tarapac\'{a}, Casilla 7D, Arica, Chile}
\affiliation{$^4$ School of Physics and Electronic-Electrical Engineering, 
	Xiangnan University, Chenzhou 423000, China}
\affiliation{$^5$ School of Physics and Optoelectronic Engineering, Foshan 
	University, Foshan 528000, China}

\date{\today }

\begin{abstract}
We investigate the formation, stability, and dynamics of solitons in a
one-dimensional binary Bose-Einstein condensate under the action of the
spin-orbit-coupling (SOC) and Lee-Huang-Yang (LHY) correction to the
underlying system of the Gross-Pitaevskii equations. We identify the
semi-dipole (SD) family of solitons and thoroughly analyze its properties.
The numerical analysis reveals intricate bifurcations, including transitions
from real to complex-valued stationary wavefunctions of the SD solitons and
norm-dependent dynamical instabilities. Stability maps in the plane of the
solitons' norm and interaction strength exhibit areas of monostability,
oscillatory behavior, and soliton splitting. Solitons with complex stationary
wavefunctions emerge as ground states in broad parameter areas, due to the
effects of the LHY terms. The other soliton species, in the form of mixed 
modes (MMs), does not feature the compexification bifurcation. In the 
LHY-dominated regime, the SD and MM solitons exhibit identical values of the 
energy for the same norm. The results deepen the understanding of nonlinear 
matter-wave states and reveal multi-stable ones in quantum gases.
\end{abstract}

\maketitle

%\preprint{}




%\thanks{}


%\keywords{}



\section{Introduction}

Spin-orbit coupling (SOC), initially explored in solid-state systems, has
become a pivotal tool in ultracold quantum gases, significantly enriching
their phenomenology \cite{Galitski2013,Zhang2016,Zhai2015}. Recent
experimental realizations of synthetic SOC in ultracold Bose-Einstein
condensates (BECs), using the Raman laser illumination, have paved the way
for exploring various quantum phenomena, including novel quantum phases,
spin textures, and topological excitations \cite%
{Lin2011,Ji2014,Goldman2014,Dalibard2011}. Complementing these experiments,
theoretical studies have extensively investigated stationary states and
nonlinear dynamics emerging from the interplay of SOC, mean-field (MF)
self-interactions, and beyond-MF effects induced by quantum fluctuations
\cite{Sakaguchi2014,Malomed2016,Li2018,Sakaguchi2018,Tononi2019}.

Solitons represent one of the fundamental species of nonlinear excitations
in ultracold gases, exhibiting stability and coherence due to the balance
between the quantum pressure and nonlinearity \cite%
{Kevrekidis2015,Khaykovich2002,Strecker2002}. Bright solitons, extensively
studied in the framework of scalar and spinor BECs, serve as versatile
testbeds to explore nonlinear dynamics in quantum matter \cite%
{Nguyen2014,Kartashov2019,Malomed2005}. Incorporating SOC into BEC systems
further diversifies the soliton phenomenology, giving rise to novel species
of self-trapped states, such as semi-vortex, semi-dipole (SD), and
mixed-mode (MM) solitons, which were primarily explored in multi-dimensional
setups \cite%
{Sakaguchi2014a,Kartashov2013,Kartashov2014,Lobanov2014,Gautam2017,Xu2013}.

In one-dimensional (1D) binary SOC-BEC systems, earlier theoretical studies
were chiefly focused on soliton solutions and their dynamics under the
action of MF interactions \cite%
{Kartashov2013,Achilleos2013,Sakaguchi2014b,Merkl2010}. However, recent
studies emphasize that effects induced by quantum fluctuations around MF
states play a crucial role, especially near the boundary of the MF
stability, significantly affecting the soliton formation, robustness, and
phase transitions \cite{Petrov2016,Astrakharchik2018,Otajonov2019,Tylutki}.
The Lee-Huang-Yang (LHY) correction, which encapsulates the averaged effect
of quantum fluctuations, has been demonstrated to stabilize self-bound
quantum droplets, which were observed in binary \cite%
{Cabrera2018,Semeghini2018,DErrico2019} and dipolar \cite%
{Ferrier2016,Schmitt2016,Baillie2016,Bottcher2021} atomic BECs . Recent
theoretical investigations also highlight significant modifications in the
soliton stability and dynamics induced by the LHY corrections in SOC BECs
\cite{Gangwar2022,Cheiney2018,Cui2021,Cappellaro2018}.

Motivated by these developments, our work aims to systematically investigate
the existence, stability, and dynamical properties of solitons in the 1D
binary BEC system under the combined action of SOC and LHY corrections.
We identify and thoroughly analyze the semi-dipole (SD) soliton family,
exploring its stability, bifurcations, and complex dynamics. In particular, 
the analysis uncovers novel bifurcations, which lead from real to complex 
soliton wavefunctions, following the variation of the nonlinearity strength. 
The results highlight the crucial role of the LHY effect and SOC in the 
emergence and stabilization of the novel soliton states.

This subsequent presentation is organized as follows. The theoretical model
is introduced in Section II. Then, Section III reports systematically
produced numerical findings which reveal the stability, structural
transitions, and dynamics of the solitons. Finally, Section IV summarizes
the results and discusses their implications for extension of theoretical
and experimental studies.

\section{The model}


We consider a homogeneous binary Bose gas confined to one spatial dimension,
composed of two components with identical masses $m$ and densities $n_{1,2}$%
, which represent different hyperfine states of the same atomic species. The
interatomic interactions are modeled by contact pseudopotentials, so that
both components feature the MF self-repulsion with a common strength $%
g_{11}=g_{22}\equiv g>0$, while the inter-component attraction is
represented by coupling $g_{12}<0$. The system's scaled energy density,
derived from the MF theory and amended by the LHY correction \cite{Lee1957},
which accounts for the effect of quantum fluctuations around the MF state,
is \cite{Petrov2016}

\begin{equation}
\begin{aligned}
E_{\mathrm{1D}}\left( n_{1},n_{2}\right) =&\frac{g}{2}\left(
n_{1}-n_{2}\right) ^{2}+\frac{\delta g}{4}\left( n_{1}+n_{2}\right) ^{2} \\
&-\frac{2}{3\pi }g^{3/2}\left( n_{1}+n_{2}\right) ^{3/2},  \label{eq:E1D}
\end{aligned}
\end{equation}%
with $\delta g\equiv g+g_{12}$. We address the binary system in the region
of weak overall MF repulsion, $0\leq \delta g\ll g$. In this case, when the
MF intra- and inter-component interactions nearly cancel each other, the LHY
corrections is a significant term. Note that it gives rise to effective
self-attraction in the 1D limit (on the contrary to the repulsion in the
multi-dimensional case \cite{Petrov2015,Petrov2016}).

The binary system is represented by the two-component MF wave function, $%
\left\{ \psi _{1}(x),\psi _{2}(x)\right\} $, which determines the respective
densities, $n_{1,2}=|\psi _{1,2}|^{2}$, and the energy functional,

\begin{equation}
\begin{aligned}
\mathscr{E}=&\int_{-\infty }^{+\infty }dx\bigg\{ E_{1\mathrm{D}}\Bigl(
\left\vert \psi _{1}\right\vert ^{2},\left\vert \psi _{2}\right\vert
^{2}\Bigr) \\
&+\sum_{j=1,2}\Bigl[ \frac{1}{2}\left\vert \partial _{x}\psi
_{j}\right\vert ^{2}-(-1)^{j}\gamma \psi _{j}^{\ast }\partial _{x}\psi _{3-j}%
\Bigr] \bigg\}.
\end{aligned} \label{eq:den_eng}
\end{equation}%
In addition to the basic energy density (\ref{eq:E1D}), it includes the
gradient (kinetic) energy of each component and the effective SOC with a
real coefficient $\gamma$. This low-energy model reliably predicts static
configurations in the above-mentioned range, $0\leq \delta g\ll g$ \cite%
{Astrakharchik2018}.

Throughout this work, we adopt natural units by setting $\hbar = m = 1$, 
so that all physical quantities are expressed in the dimensionless form.

The variational procedure applied to the energy functional (\ref{eq:den_eng}) 
produces a system of scaled coupled Gross-Pitaevskii equations (GPEs) \cite%
{Gangwar2022}:

\begin{equation}
\begin{aligned}
i\partial _{t}\psi _{j}=&\Bigl[ -\frac{1}{2}\partial _{xx}+\frac{\delta g}{2}%
\left( \left\vert \psi _{1}\right\vert ^{2}+\left\vert \psi
_{2}\right\vert ^{2}\right) \\
&-(-1)^{j}g\Bigl( \left\vert \psi _{1}\right\vert^{2}-\left\vert \psi _{2}%
\right\vert ^{2}\Bigr) \\
& -\frac{g^{3/2}}{\pi }\left( \left\vert \psi _{1}\right\vert^{2}+\left\vert %
\psi _{2}\right\vert ^{2}\right) ^{1/2}\Bigr] \psi _{j} \\
& -(-1)^{j}\partial _{x}\psi _{3-j},~~j=1,2,
\end{aligned}  \label{eq:GPEs_1}
\end{equation}%
where the SOC coefficient is fixed to be $\gamma \equiv 1$ by means of
rescaling. Dynamical invariants of this system are the energy, scaled number
of atoms (alias the total norm of the wave function),%
\begin{equation}
N=\int_{-\infty }^{+\infty }(|\psi _{1}|^{2}+|\psi _{2}|^{2})dx\equiv
N_{1}+N_{2},  \label{N}
\end{equation}%
and the total momentum,
\begin{equation}
P=i\int_{-\infty }^{+\infty }\left( \psi _{1}\frac{\partial }{\partial x}%
\psi _{1}^{\ast }+\psi _{2}\frac{\partial }{\partial x}\psi _{2}^{\ast
}\right) dx.
\end{equation}
The overall characteristic of the two-component solitons determines the
asymmetry between the norms of its components:%
\begin{equation}
	\eta =(N_{1}-N_{2})/(N_{1}+N_{2}),  \label{eta}
\end{equation}%
where $N_1$ and $N_2$ follow the definitions adopted in Eq.~(\ref{N}).

Note that, due to the use of natural units ($\hbar=m=1$) and several 
rescalings applied for the simplification of the coupled Gross–Pitaevskii 
equations, the scaled norm $N$ defined in Eq. (\ref{N}) does not directly 
correspond to the number of atoms. Based on the scaling relations and 
comparison with previous works \cite{Petrov2016,Astrakharchik2018}, we 
estimate typical values of the atom number $N_{\textrm{phys}}$ corresponding 
to our results. For instance, using characteristic values of the scattering 
length $a\sim100,\text{nm}$ and transverse confinement length $a_{\perp}%
\sim1-5\,\mu \mathrm {m}$, the relation $N_{\mathrm{phys}}\sim N \times(%
a_{\perp}/a)$ suggests that the scaled norms used in this work correspond 
to $N_{\textrm{phys}}\sim10^3–10^4$, which certainly belongs to the 
experimentally achievable ranges in ultracold atomic gases.

Our first objective is to construct 1D solitons as solutions of Eqs.~(\ref%
{eq:GPEs_1}). In the absence of the LHY terms, 1D two-component solitons
supported by SOC were investigated previously \cite%
{Achilleos2013,Kartashov2013}. We here focus on the opposite case of the 
LHY superfluid, with $\delta g=0$, when the dominant nonlinearity is 
represented solely by the LHY terms~\cite{Jorgensen2018}. In this case, 
Eq.~(\ref{eq:GPEs_1}) simplifies to the form, in which the mean-field 
interaction term $\sim\delta g$ vanishes and the nonlinearity amounts to 
the LHY correction. Assuming stationary solutions of the form $\psi_{1,2}%
= e^{-i\mu t}\phi_{1,2}(x)$ and substituting them into the reduced equations 
yields the coupled stationary equations

\begin{equation}
\begin{aligned}
\mu \phi _{1}=&-\frac{1}{2}\phi _{1}^{\prime \prime }+\phi _{2}^{\prime
}+g(|\phi _{1}|^{2}-|\phi _{2}|^{2})\phi _{1} \\
& -\frac{g^{3/2}}{\pi } (|\phi_{1}|^{2}+|\phi _{2}|^{2})^{1/2}\phi _{1},
\end{aligned} \label{eq:phi1}
\end{equation}%
\begin{equation}
	\begin{split}
		\mu \phi _{2}=&-\frac{1}{2}\phi _{2}^{\prime \prime }-\phi _{1}^{\prime
		}+g(|\phi _{2}|^{2}-|\phi _{1}|^{2})\phi _{2} \\
		& -\frac{g^{3/2}}{\pi }(|\phi_{1}|^{2}+|\phi _{2}|^{2})^{1/2}\phi _{2},
	\end{split} \label{eq:phi2}
\end{equation}%
where the prime stands for $d/dx$. These equations admit both real and
complex solutions, as shown below.

\section{Results}

In this section, we present numerical results for the variety of soliton
solutions of the simplified Eq.~(\ref{eq:GPEs_1}) in the LHY-only regime, 
obtained by means of the accelerated imaginary time evolution method (AITEM). 
The stability of the solitons was then tested by adding random noise, with 
the relative amplitude at the $1\%$ level, to the solutions of the simplified 
Eq.~(\ref{eq:GPEs_1}) in the LHY-only regime and performing direct simulations. 
Our findings reveal that the system gives rise to two distinct species of 
solitons, namely SD and MM ones, which are generated by different inputs. 
The primary focus of the study is the effect of the total norm on the 
structure of the solitons and phase transitions between them.

\subsection{SD\ (semi-dipole) solitons}

First, we employed AITEM to solve the coupled GPEs, initiating the
imaginary-time integration with the following input:

\begin{eqnarray}
\psi _{1}(t &=&0)=A_{1}\exp (-x^{2}/l_{1}^{2}),  \label{SD1} \\
\psi _{2}(t &=&0)=A_{2}x\exp (-x^{2}/l_{2}^{2}),  \label{SD2}
\end{eqnarray}%
including the spatially even and odd components $\psi _{1}(x)$ and $\psi
_{2}(x)$, respectively, with amplitudes $A_{1}=1$, $A_{2}=0.5$ and common
width $l_{0}=\sqrt{10}\approx \allowbreak 3.16$. The established SD modes
keep the same parities of their components. By way of this approach, we
identify two distinct types of SD solitons, represented by purely real and
complex stationary wavefunctions, respectively. The SD solitons of the former
type are obtained with a relatively small norm. A transition to solitons with
complex stationary wavefunctions occurs with the increases of the norm.

\begin{figure}[tbp]
\includegraphics[width=6.5cm]{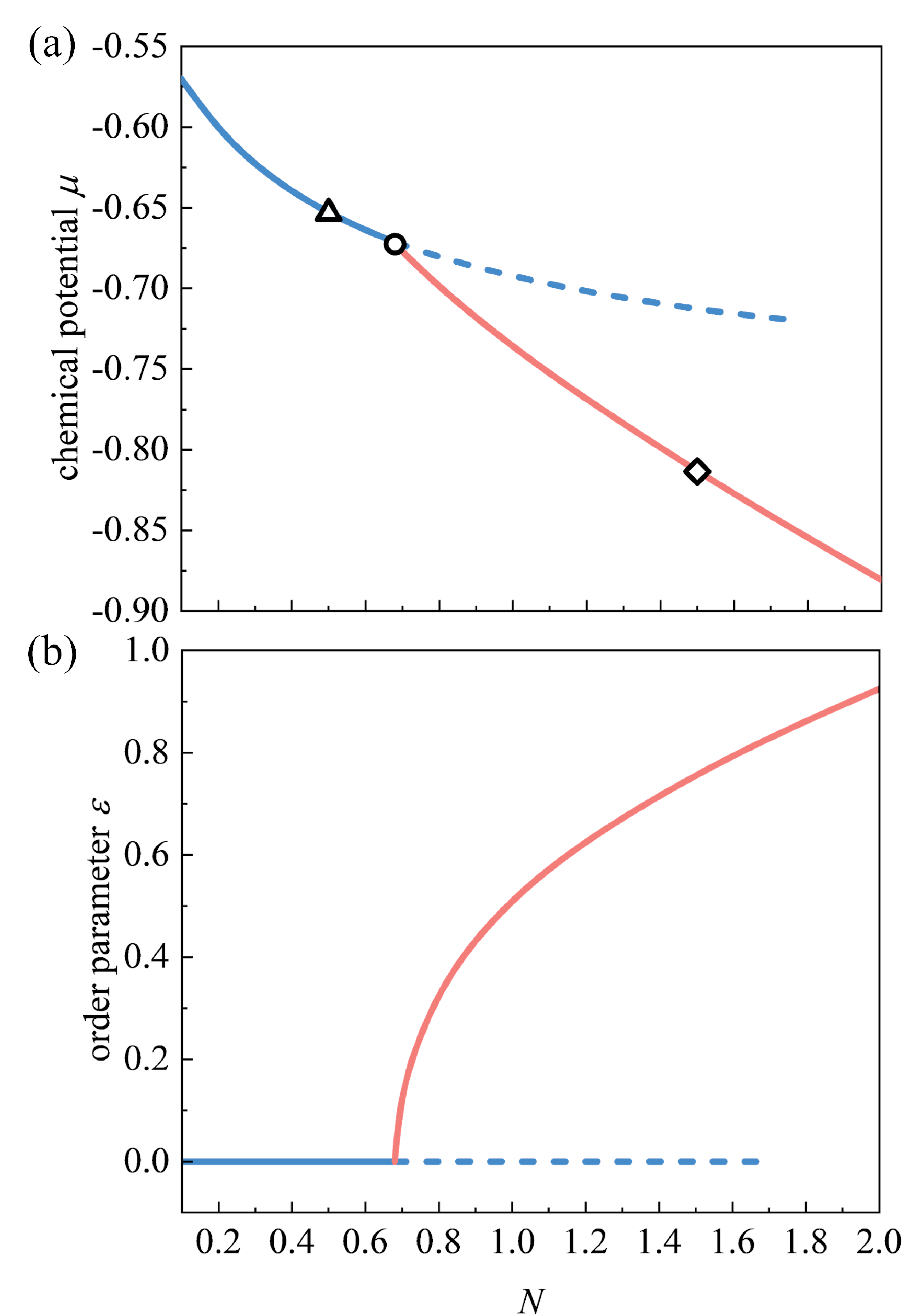}
\caption{(Color online) (a) The chemical potential of the two types of the
SD (semi-dipole) solitons as a function of their norm $N$, featuring the
bifurcation at $N=N_{\text{cr}}^{(1)}\approx 0.67$ (marked by the circle),
for the characteristic value of the interaction constant, $g=\protect\pi %
^{2/3}$, in the reduced form of Eq.~(\ref{eq:GPEs_1}) for $\delta g=0$. 
The family of the solitons with real stationary wavefunction, is stable 
(being represented by the solid blue curve) below the bifurcation point, 
at $N<N_{\text{cr}}^{(1)}$, and unstable (shown by the dashed blue curve) 
above the bifurcation, at $N>N_{\text{cr}}^{(1)}$. The family of the SD 
solitons with complex wavefunctions (plotted by the red line), exists and 
is stable at $N>N_{\text{cr}}^{(1)}$. (b) The forward (supercritical) character 
of the bifurcation is exhibited by the dependence of the order parameter 
(\protect\ref{eps}) on $N$. The blue and red curves again represent the SD 
solitons with real and complex wavefunctions, respectively.} 
\label{fig:CP&EPS_SD}
\end{figure}

\begin{figure}[tbp]
\includegraphics[width=6.5cm]{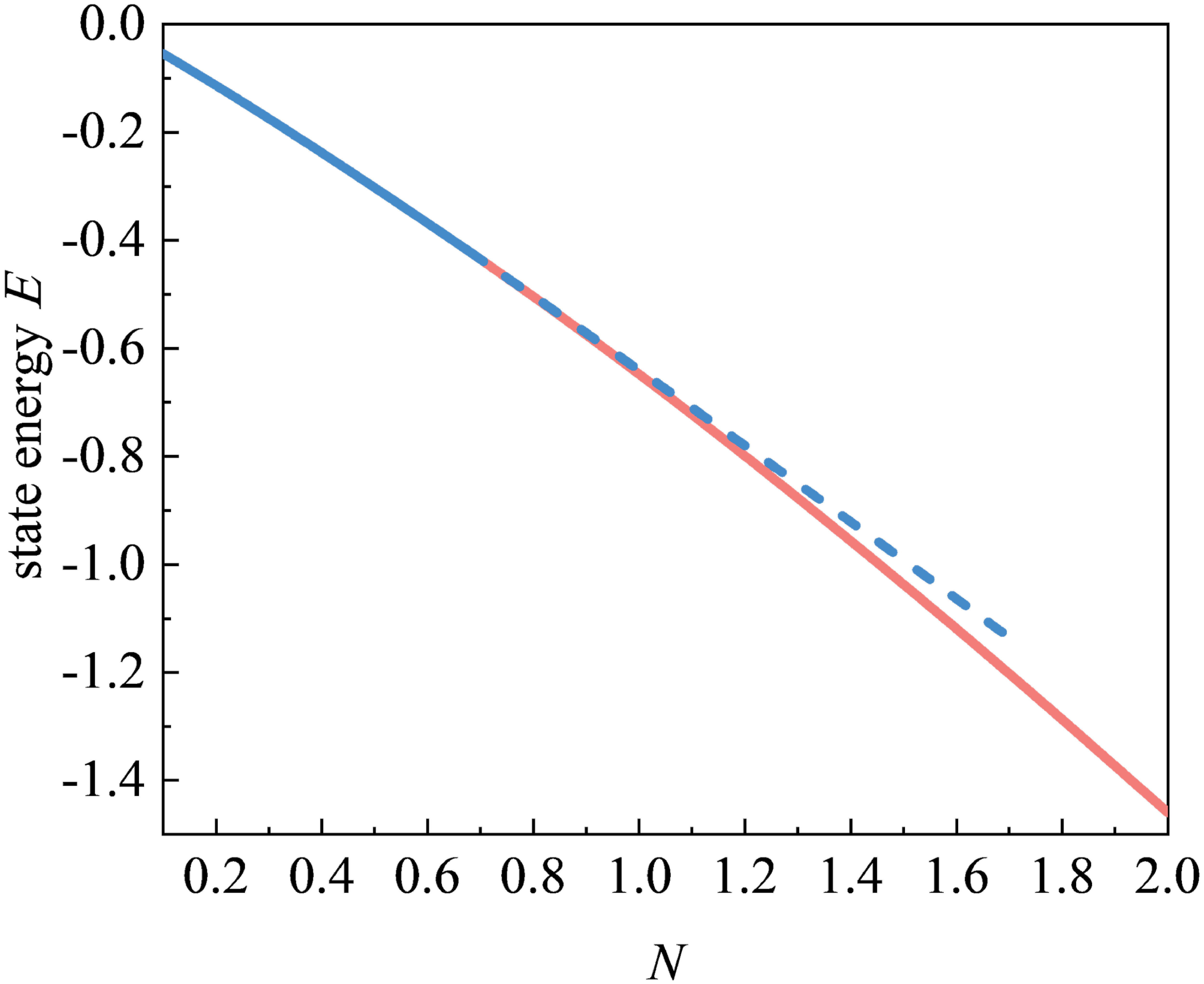}
\caption{(Color online) The total energy of two types of the SD solitons as
a function of $N $, for $g=\protect\pi ^{2/3}$. Above the bifurcation point
(at $N>N_{\text{cr}}^{(1)}$), the soliton with the complex wavefunction,
plotted by the red line, realizes the system's GS with the minimal energy,
while the family of the solitons with the real wavefunction (plotted by the
blue curve) has a larger energy, being unstable.} \label{fig:E_SD}
\end{figure}

As a representative case, we take the contact-interaction parameter in the 
simplified Eq.~(\ref{eq:GPEs_1}) in the LHY-only regime as $g=\pi ^{2/3}$. 
The analysis reveals a well-defined phase transition (bifurcation) between 
the real and complex types of the SD solitons, as their norm $N$ increases. 
Figure~\ref{fig:CP&EPS_SD}(a) displays the bifurcation by plotting the 
chemical potential $\mu $ of the SD solitons vs. $N$. Note that the 
dependence $\mu (N)$ for both the real and complex solitons satisfies the 
Vakhitov--Kolokolov criterion, $d\mu /dN<0$, which is the commonly known 
necessary stability condition for solitons of the nonlinear-Schr\"{o}dinger 
type~\cite{VakhKol,Kuznetsov2011}. The SD family with the real stationary 
wavefunction, represented by the blue lines, remains stable (the solid 
segment of the blue lines) below the bifurcation point, i.e., at 
$N<N_{\text{cr}}^{(1)}\approx 0.67$ in Fig.~\ref{fig:CP&EPS_SD}(a), and 
extends as an unstable one to $N>N_{\text{cr}}^{(1)}$. The branch of the SD 
solitons with the complex stationary wavefunctions, represented by the red 
lines, emerges as a stable one above the bifurcation point, i.e., at 
$N>N_{\text{cr}}^{(1)}$. When this branch exists, it represents the system's 
ground state (GS) with the minimal energy, as shown in Fig.~\ref{fig:E_SD}, 
where the energy of both types of the SD solitons is plotted vs. $N$. 
From the experimental perspective, the bifurcation point at 
$N_{\text{cr}}^{(1)} \approx 0.67$, according to the estimate given 
in Sec.~II, corresponds to a few thousand atoms, which implies an experimentally 
accessible regime with atomic species such as $^{7}$Li or $^{39}$K under 
strong transverse confinement. We have verified that stationary 
states with other parity setups (in particular, with both components even 
or odd) for the same norm have positive total energy, therefore they are 
not self-trapped localized modes.

To characterize the bifurcation (phase transition) of the solitons, we
define the order parameter, $\varepsilon $, as the square root of the total
norm of the imaginary components in the binary wavefunctions, i.e.,

\begin{equation}
\varepsilon =\left[ \int_{-\infty }^{+\infty }\left( \left\vert \frac{\psi
_{1}-\psi _{1}^{\ast }}{2}\right\vert ^{2}+\left\vert \frac{\psi _{2}-\psi
_{2}^{\ast }}{2}\right\vert ^{2}\right) dx\right] ^{1/2}.  \label{eps}
\end{equation}%
Figure~\ref{fig:CP&EPS_SD}(b) reveals the relationship between $\varepsilon $
and $N$. The bifurcation, which takes place at $N=N_{\text{cr}}^{(1)}$,
is of the forward (supercritical) type \cite{Iooss1980}, alias the phase
transition of the second kind. Accordingly, the SD solitons with the real
wavefunctions are stable at $N<N_{\text{cr}}^{(1)}$, and become unstable
at $N>N_{\text{cr}}^{(1)}$, where the \emph{forward-going} branches of
stable solitons with the complex wavefunction are stable (hence the name
of the forward bifurcation).

It should be noted that, for SD-type solitons in the system
with the interaction constant $g=\pi^{2/3}$, the norms $N_1$ and $N_2$ remain
virtually identical (with differences $\sim 10^{-3}$), confirming the specific 
nature of the SD states for this value of the interaction strength. In this 
connection, it is relevant to mention that relation $N_{1}\approx N_{2}$ 
ensues from Eqs.~(\ref{eq:phi1}) and (\ref{eq:phi2}) in the long-wave limit: 
if the terms with the spatial derivatives may be neglected, the equations 
reduce to the algebraic ones:%
\begin{equation}
\mu =g\left( \left\vert \phi _{1,2}\right\vert ^{2}-\left\vert \phi
_{2,1}\right\vert ^{2}\right) -\frac{g^{3/2}}{\pi }\left( \left\vert \phi
_{1}\right\vert ^{2}+\left\vert \phi _{2}\right\vert ^{2}\right) ^{1/2},
\end{equation}%
from where $N_{1}=N_{2}$ follows immediately. For comparison, the limit of
the LHY superfluid, which is considered below, being based on Eqs.~(\ref%
{Phi1}) and (\ref{Phi2}), that do not include the terms $\sim \bigl(
\left\vert \phi _{1,2}\right\vert ^{2}-\left\vert \phi _{2,1}\right\vert
^{2}\bigr) $, gives rise to states with $N_{1}\neq N_{2}$, i.e., $\eta \neq
0$ (see Eq.~(\ref{eta}) and Fig.~\ref{fig:eta_LHY} below).

Our comprehensive numerical analysis reveals that the near-equality $N_1%
\approx N_2$ [i.e., $\eta \approx 0$, see Eq.~\ref{eta}] indeed takes place
only in the long-wave limit ($g<2$), where spatial derivatives become 
negligible, enforcing the component symmetry, as argued above. For larger 
interaction strengths ($g>2$), both real and complex families of SD solitons 
exhibit significant asymmetry ($\eta\ne0$), as quantified in Fig.~\ref{fig:%
eta_SD}. Specifically, the departure from the long-wave approximation allows 
the gradient terms to amplify the norm asymmetry, which is consistent with 
the behavior observed in the LHY-dominated regime ($g \rightarrow \infty$), 
where $\eta$ attains large values, see Fig.~\ref{fig:eta_LHY} below. 

\begin{figure}[tbp]
\includegraphics[width=6.5cm]{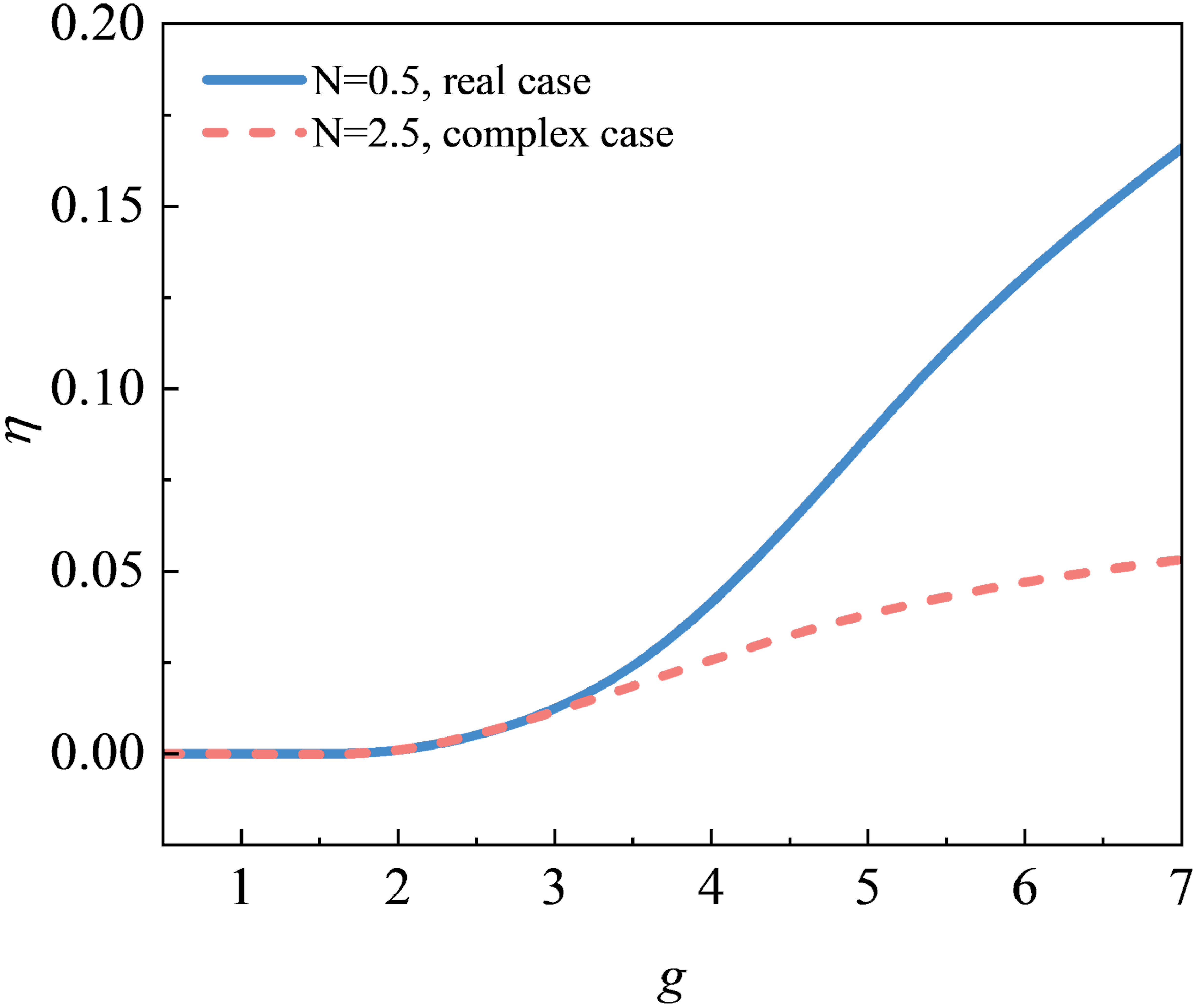}
\caption{(Color online) The norm-asymmetry parameter $\eta$ (\ref{eta}) for 
SD solitons as a function of interaction strength $g$. Results are shown for 
two characteristic norms: $N=0.5$ (soliton solutions with real wavefunctions, 
solid blue line) and $N=2.5$ (soliton solutions with dashed red line). For 
$g<2$ (long-wave limit), $\eta\approx0$ confirms near-exact equality $N_1=N_2$ 
enforced by component symmetry. For $g>2$, both soliton families exhibit 
progressive asymmetry ($\eta>0$) due to nonlinearity-driven symmetry breaking, 
with deviations exceeding numerical error bounds. The saturation of $\eta$ at 
large $g$ aligns with the LHY-superfluid asymptotics in Fig.~\ref{fig:eta_LHY}.}
\label{fig:eta_SD}
\end{figure}

\begin{figure*}[tbp]
\includegraphics[width=16.5cm]{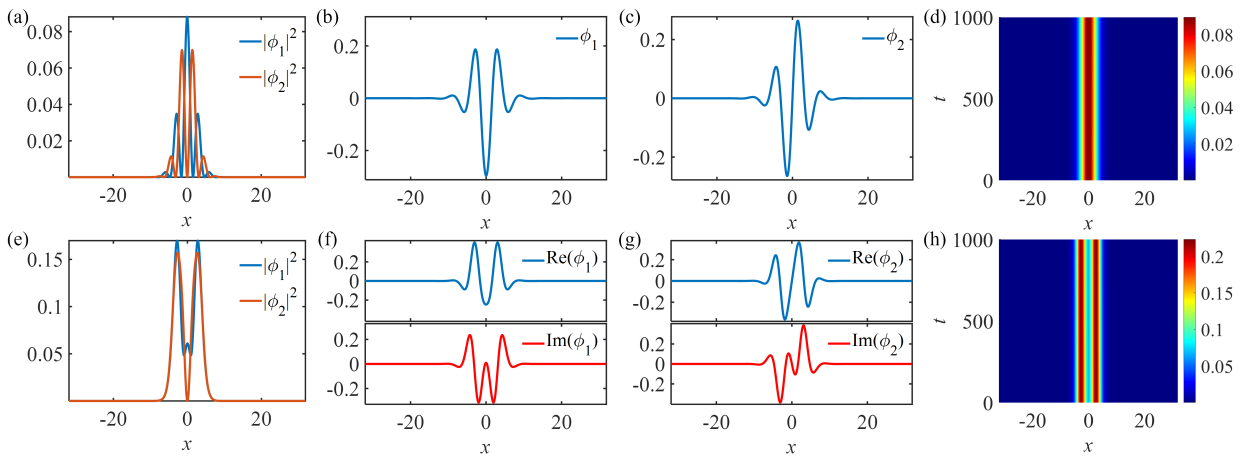}
\caption{(Color online) Stable SD solitons (1D counterparts of 2D
semi-vortex solitons) produced by the numerical solution of the simplified 
Eq.~(\ref{eq:GPEs_1}) in the LHY-only regime with $g=\protect\pi ^{2/3}$. 
Panels (a-d): A low-norm soliton [$N=0.5$, marked by the triangle in Fig.~% 
\protect\ref{fig:CP&EPS_SD}(a)]. (a) The density profile showing the 
symmetric two-lobe structure. (b) The real stationary wavefunction of the 
first component, with the even parity. (c) The real stationary wavefunction 
of the second component, with the odd-parity. (d) Real-time 
evolution of the soliton under 1\% random perturbations applied to the 
initial wavefunctions, illustrating stability or instability-induced 
dynamics. Panels (e)-(h): A high-norm soliton [$N=1.5$, marked by the 
rhombus in Fig.~\protect\ref{fig:CP&EPS_SD}(a)]. (e) The density distribution, 
featuring tighter nonlinear confinement. (f) The complex stationary 
wavefunction of the first component, with the even parity. (g) The complex 
stationary wavefunction of the second component, with the odd parity. 
(h) Real-time evolution of the soliton under 1\% random 
perturbations applied to the initial wavefunctions, revealing robustness 
against perturbations or, when unstable, the nature of the ensuing dynamics.} 
\label{fig:2StaSD}
\end{figure*}

Figure~\ref{fig:2StaSD} displays families of the stable real and complex SD
solitons. These self-trapped modes, with the opposite parities of their
components, are 1D analogs of the 2D \textit{semi-vortex solitons}, which
are well known in the 2D nonlinear SOC system \cite{Sakaguchi2014}. The
semi-vortices carry vorticities $0$ and $1$ in their coupled components. As
seen in Figs.~\ref{fig:2StaSD}(d) and (h), the solitons with real and
complex stationary wavefunctions feature, respectively, single and double
stripes in terms of the total density, $|\psi _{1}|^{2}+|\psi _{2}|^{2}$.

Panels (a)-(d) in Figs.~\ref{fig:2StaSD} depict a low-norm soliton ($N=0.5$), 
corresponding to the triangular marker in Fig.~\ref{fig:CP&EPS_SD}(a).
The density profile in (a) reveals a symmetric two-lobe structure,
consistent with the stationary real two-component wavefunctions shown in (b)
and (c). The real-time evolution in (d) is obtained by 
propagating the initial stationary wavefunctions with 1\% random perturbations, 
which serves to examine the soliton’s robustness; the results confirm its 
stability throughout the simulation time. Panels (e)-(h) display a high-norm 
soliton ($N=1.5$), corresponding to the rhombus marker in Fig.~\ref{fig:%
CP&EPS_SD}(a), where stronger nonlinearity gives rise to the SD solitons 
with the complex stationary wavefunction. The density distribution in panel 
(e) exhibits tighter self-trapping, in comparison to the low-norm soliton. 
The temporal evolution in panel (h), likewise simulated with 
1\% random noise added to the initial condition, further demonstrates the 
full stability of this soliton. The characteristic flat-top shape of the 
1D quantum droplets, which is maintained by the balance of the MF self-%
repulsion, with strength $\sim\delta g>0$ in Eq.~(\ref{eq:GPEs_1}), and 
self-attraction provided by the LHY terms \cite{Astrakharchik2018} is not 
observed here, as we have set $\delta g=0$.

The transition from real to complex stationary wavefunctions, which follows
the increase of the norm, represents a fundamental symmetry-breaking
mechanism driven by the interplay between the nonlinearity and SOC. It
distinguishes the SD solitons reported here from the previously known ones,
cf. Refs. \cite{Kartashov2013,Xu2013,Achilleos2013}.

The complexification bifurcation (phase transition) implies that a real
solution for $\phi _{1,2}(x)$ carries over into
\begin{equation}
\left( \phi _{1,2}\right) _{\mathrm{complex}}=\phi _{1,2}(x)+i\chi _{1,2}(x),
\end{equation}%
with infinitesimal real functions $\chi _{1,2}(x)$ satisfying the system of
linearized equations:
\begin{equation}
\begin{aligned}
\mu \chi _{1}=&-\frac{1}{2}\chi _{1}^{\prime \prime }+\chi _{2}^{\prime}%
+g(\phi _{1}^{2}-\phi _{2}^{2})\chi _{1} \\
&-\frac{g^{3/2}}{\pi }(\phi_{1}^{2}+\phi _{2}^{2})^{1/2}\chi _{1},
\end{aligned} \label{eq:chi1}
\end{equation}
\begin{equation}
\begin{aligned}
\mu \chi _{2}=&-\frac{1}{2}\chi _{2}^{\prime \prime }-\chi _{1}^{\prime}%
+g(\phi _{2}^{2}-\phi _{1}^{2})\chi _{2} \\
&-\frac{g^{3/2}}{\pi }(\phi_{1}^{2}+\phi _{2}^{2})^{1/2}\chi _{2}.  
\end{aligned} \label{eq:chi2}
\end{equation}%
An obvious (actually, trivial) solution of Eqs.~(\ref{eq:chi1}) and (\ref%
{eq:chi2}) is $\chi _{1,2}(x)\equiv \phi _{1,2}(x)$, which corresponds to
the fact that solutions of the simplified Eq.~(\ref{eq:GPEs_1}) in the LHY-%
only regime are invariant with respect to the phase shift [multiplication of 
both $\psi_{1,2}$ by $\exp(i\theta)$ with an infinitesimal phase $\theta $].

As a formal generalization of Eqs.~(\ref{eq:chi1}) and 
(\ref{eq:chi2}), we also consider the corresponding linear eigenvalue problem 
in which chemical potential $\mu$ is replaced by a general eigenvalue 
$\nu$. This formulation is used to identify the onset of the complexification 
bifurcation when $\nu$ coincides with $\mu$, signalling the appearance of 
a nontrivial solution in addition to the trivial phase-rotation mode.

An obvious solution of this eigenvalue problem is $\nu =\mu$, but other 
solutions may exist too. If there is an eigenvalue $\nu \neq \mu$, the 
complexification bifurcation takes place at value $\mu_{\mathrm{crit}}$ of 
parameter $\mu $ in Eqs.~(\ref{eq:phi1}) and (\ref{eq:phi2}) at which the 
additional eigenvalue $\nu $ becomes equal to $\mu $, i.e., $\mu $ becomes 
a \emph{double eigenvalue}. Indeed, in this special case there is a solution 
of the generalized eigenvalue problem based on Eqs.~(\ref{eq:chi1}) and 
(\ref{eq:chi2}) with $\mu\to\nu$ which is different 
from the trivial one, $\chi _{1,2}(x)\equiv \phi _{1,2}(x)$, and this 
nontrivial solution will initialize the onset of the complexification at 
$\mu =\mu _{\mathrm{crit}}$.

\begin{figure}[tbp]
\includegraphics[width=6.5cm]{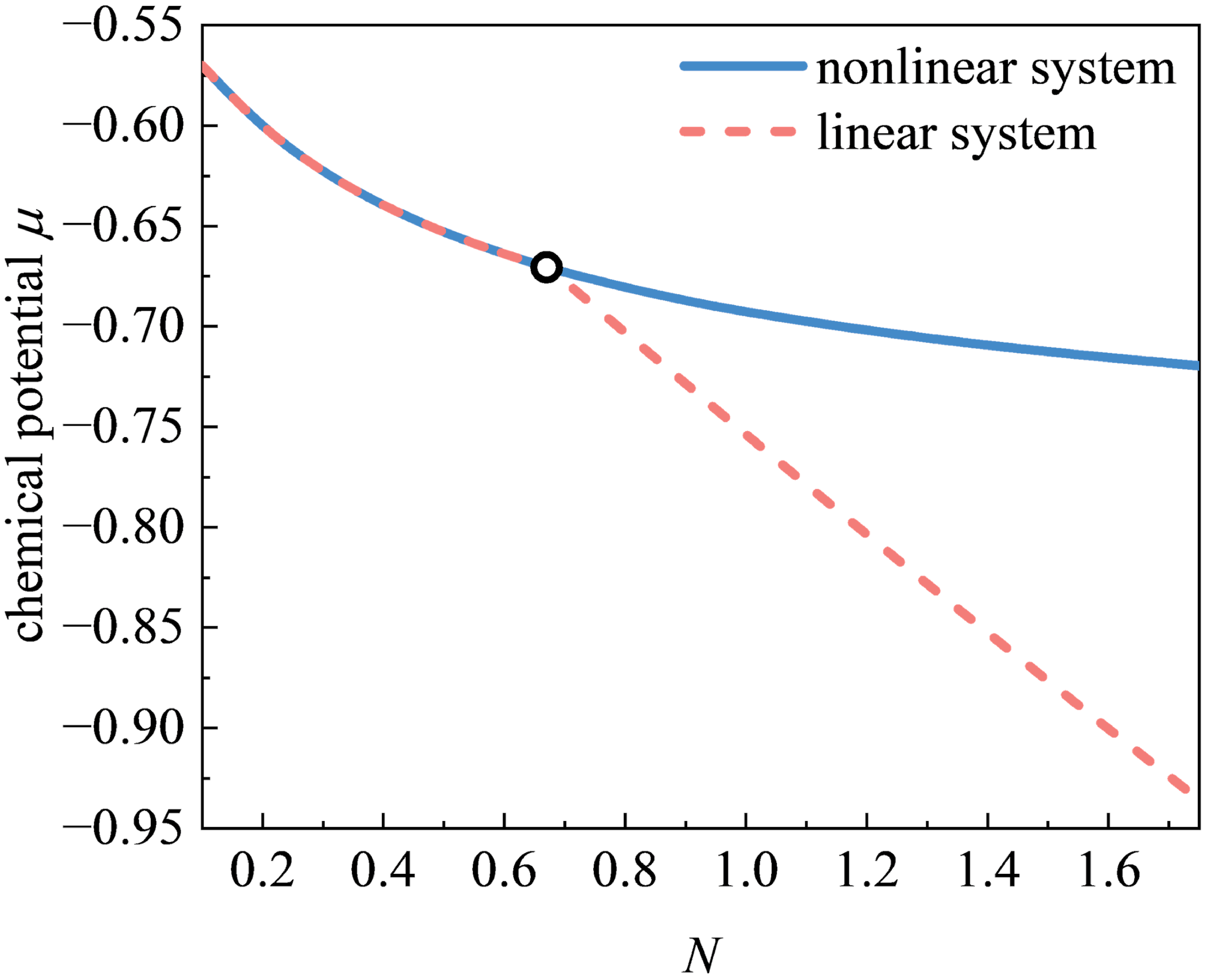}
\caption{(Color online) The chemical potential of the soliton solutions
produced by the numerical solution of the nonlinear and linear systems [Eqs.~%
(\protect\ref{eq:phi1}), (\protect\ref{eq:phi2}) and (\protect\ref{eq:chi1}),
(\protect\ref{eq:chi2}), respectively, with $g=\protect\pi ^{2/3}$]. The
blue solid and red dashed curves correspond, severally, to the nonlinear and
linear system. The bifurcation point ($N_{\text{cr}}\approx 0.67$) separates
two distinct regimes: (i) Below the point (at $N<N_{\text{cr}}$), both systems
yield identical chemical potentials; (ii) Above the bifurcation (at $N>N_{%
\text{cr}}$), the chemical potential of the linear system exhibits a linear
dependence on norm $N$ (the dashed curve), whereas the nonlinear system
deviates from this trend (the solid curve). These results fully agree with
those displayed by the blue curve in Fig.~\protect\ref{fig:CP&EPS_SD}(a). }
\label{fig:mu_Nonlinear&Linear}%

\end{figure}

To validate this hypothesis, we numerically solved the system of Eqs.~(\ref%
{eq:phi1}) and (\ref{eq:phi2}), obtaining wavefunctions $\phi_{1}$, $\phi_{2}$, 
and the corresponding chemical potentials. Subsequently, these wavefunctions 
were substituted into the generalized linear eigenvalue problem based on 
Eqs.~(\ref{eq:chi1}) and (\ref{eq:chi2}), in which $\mu$ is replaced by 
a free eigenvalue $\nu$; the resulting system was then solved numerically 
to determine $\chi_1$, $\chi_2$, and the corresponding eigenvalue $\nu$. 
Figure~\ref{fig:mu_Nonlinear&Linear} displays the evolution of the chemical 
potential of the soliton solutions for both the nonlinear (the blue solid 
curve) and linear (the red dashed curve) systems. The results demonstrate 
full agreement with the blue curve in Fig.~\ref{fig:CP&EPS_SD}(a). The 
bifurcation point is identified at $N_{\text{cr}}\approx 0.67$: below this 
point (at $N<N_{\text{cr}}$), the nonlinear and linear systems yield identical 
solutions, in exact agreement with the above-mentioned fact that the 
generalized linear eigenvalue problem based on Eqs.~(\ref{eq:chi1}) and 
(\ref{eq:chi2}), with $\mu$ replaced by $\nu$, admits the obvious (actually, 
trivial) solution with $\nu =\mu $. Above the bifurcation (at 
$N>N_{\text{cr}}$), however, the solutions of the two systems separate, again 
in agreement with the above-mentioned existence of the nontrivial solution 
of the linear system at $N>N_{\text{cr}}$. Note that the chemical potential 
produced by the solution of the linear system exhibits a linear dependence 
on norm $N$, whereas the nonlinear system deviates from this trend.

\begin{figure*}[tbp]
\includegraphics[width=16.5cm]{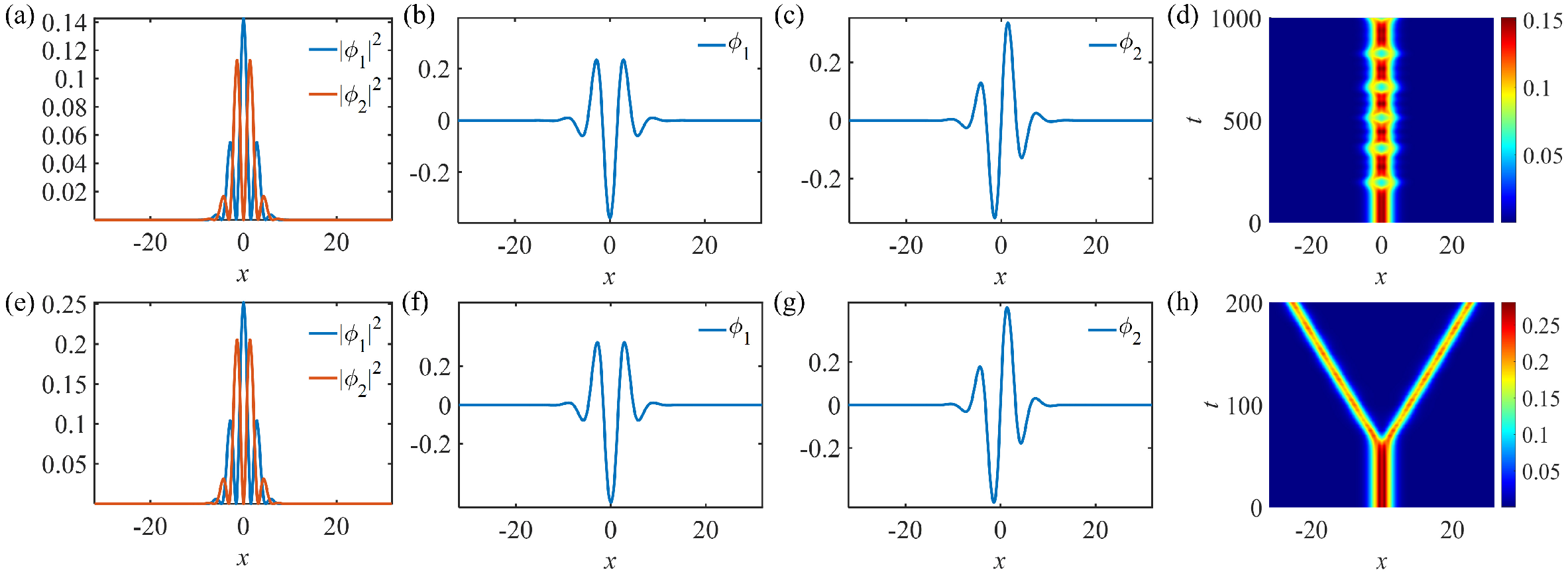}
\caption{(Color online) The instability of the SD solitons with the real
stationary wavefunctions and different norms, in the case of $g=\protect\pi %
^{2/3}$. Panels (a-d): A low-norm soliton ($N=0.8$). (a) The density
profile; (b) the first component of the real stationary wavefunction
(even-parity, purely real); (c) the second component of the real stationary
wavefunction (odd-parity, purely real). (d) The real-time 
evolution, obtained by adding 1\% random perturbations to the initial 
stationary wavefunctions, indicates weak instability of the low-norm soliton, 
which keeps its size and exhibits small-amplitude oscillations. Panels 
(e-h): A high-norm soliton ($N=1.5$). (e) The density profile similar to 
that of the low-norm soliton; (f) the first-component real stationary 
wavefunction (even-parity); (g) the second-component real stationary 
wavefunction (odd-parity). (h) The real-time evolution, 
likewise simulated with 1\% random noise in the initial condition, reveals 
strong instability, which leads to the splitting of the soliton into two 
separating fragments.} \label{fig:2UnStaSD}
\end{figure*}

In the instability region of the solution with the real stationary
wavefunction, represented by the dashed curve in Fig.~\ref{fig:CP&EPS_SD}%
(a), the dynamical behavior of the solitons is significantly affected by the
norm. Figure~\ref{fig:2UnStaSD} reports the perturbed perturbed real-time 
evolution (initiated by adding 1\% random noise to the stationary wavefunctions) 
of two unstable SD solitons. In the low-norm case [$N=0.8$, 
panels (a-d)], the unstable soliton maintains its spatial structure, exhibiting 
only weak oscillations under the applied perturbation (panel d). In this 
regime, the unstable solitons keep the phase locking between the even-parity 
first and odd-parity second components, whose unperturbed real stationary 
shapes are displayed in panels (b) and (c), respectively. In contrast to 
that, the unstable soliton with a high norm [$N=1.5$, whose stationary real 
wavefunctions are shown in panels (e-h)] demonstrates strong instability: 
the initially balanced configuration [see panels (e-g)], which features 
tighter self-trapping, in comparison to the low-norm soliton, 
breaks apart into two separating fragments during the 
perturbed evolution (panel h). The oscillation period in 
Fig.~\ref{fig:2UnStaSD}(d) and the splitting time in Fig.~\ref{fig:2UnStaSD}(h) 
provide representative values for the adopted realization of the 1\% random 
noise, and may slightly vary for different noise realizations. The norm-%
dependent transition from the oscillatory dynamics to the splitting 
highlights the role of the nonlinearity strength.

The soliton solutions are strongly affected by both norm $N$ and the
contact-interaction parameter, $g$. The above consideration addressed the
role of $N$ for fixed $g=\pi ^{2/3}$. Our findings reveal that, in the
general case, for fixed $g$ (which may be different from $g=\pi ^{2/3}$),
the increase of $N$ drives the SD solitons through three distinct
transitions, which are determined by three sequentially ordered critical
norms $N_{\text{cr}}^{(1)}<N_{\text{cr}}^{(2)}<N_{\text{cr}}^{(3)}$ at a
fixed interaction strength $g$. When the norm passes from $N<$ $N_{\text{cr}%
}^{(1)}$ to $N>N_{\text{cr}}^{(1)}$ the system undergoes a fundamental
stability transition: the soliton solutions with real wavefunctions, which
were stable at $N<N_{\text{cr}}^{(1)}$, enter a dynamically unstable regime.
Actually, the value $N_{\text{cr}}^{(1)}(g=\pi ^{2/3})\approx 0.67$ is
precisely the bifurcation point shown above in Figs.~\ref{fig:CP&EPS_SD} and %
\ref{fig:E_SD} for $g=\pi ^{2/3}$. Beyond this threshold, as $N$ further
increases past $N_{\text{cr}}^{(2)}$, the primary instability mechanism
switches from intrinsic oscillations of the solitons to their irreversible
splitting. Finally, the soliton solutions with real wavefunctions cease
to exist for $N>N_{\text{cr}}^{(3)}$. Crucially, for $N>N_{\text{cr}}^{(1)}$,
the soliton solutions with complex wavefunctions emerge as the GS, coexisting
with progressively destabilized soliton solutions with real wavefunction
until their extinction at $N_{\text{cr}}^{(3)}$.

\begin{figure}[tbp]
\includegraphics[width=6.5cm]{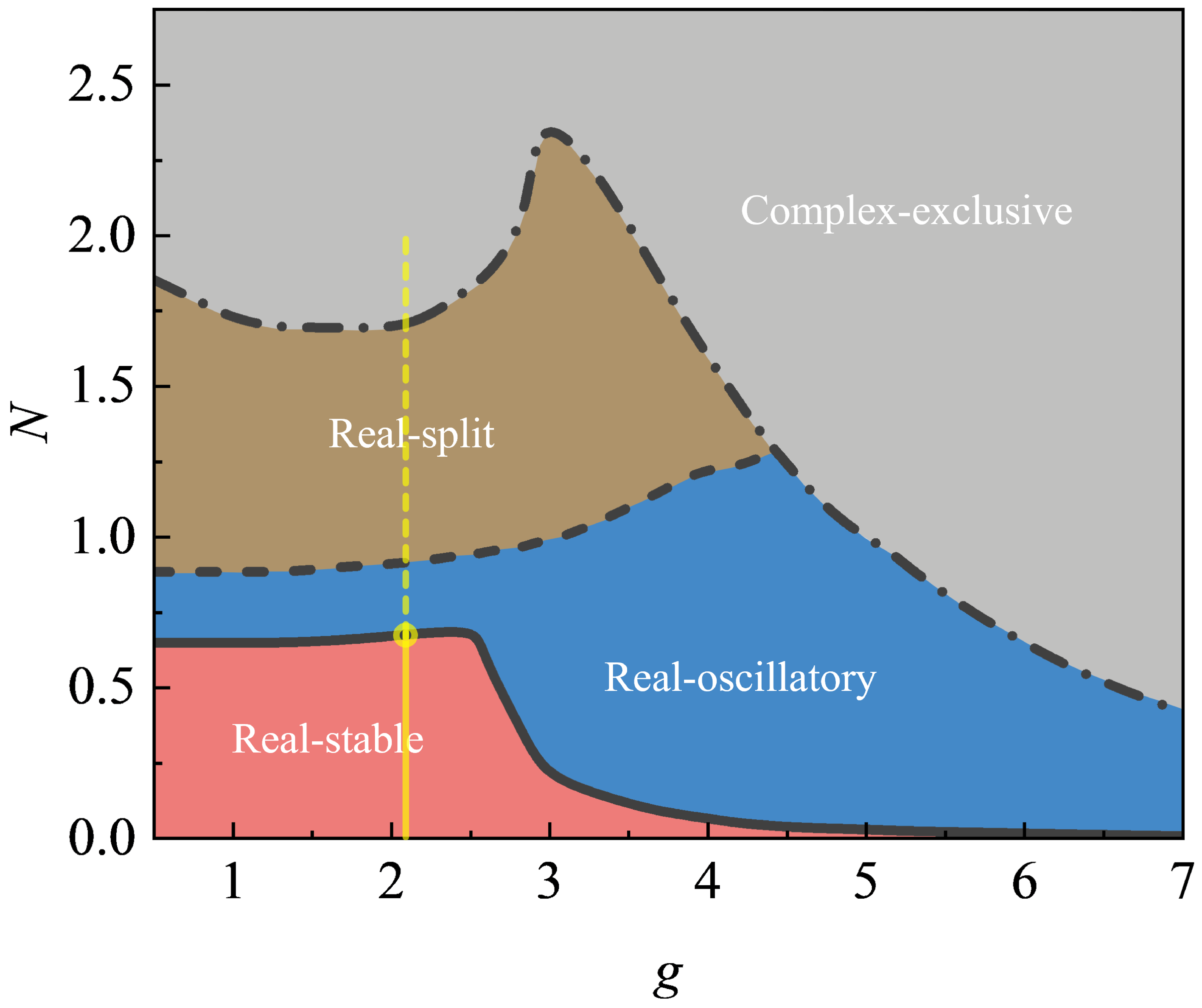}
\caption{(Color online) The dependence of three critical bifurcation
thresholds, $N_{\text{cr}}^{(1,2,3)}$, on functions of the interaction
parameter $g$. The respective parameter plane of $\left( g,N\right) $ is
partitioned in four dynamical regimes, as described in the main text:
monostable SD solitons (the red area below the solid curve); oscillatory
states (breathers, in the blue area); the soliton-splitting states (the brown
area); and the gray area above the dashed-dotted curve, where solely the
stable stationary solitons with the complex wavefunctions exist (such stable
solitons exists as well in the blue and brown areas). The yellow vertical
line, plotted at $g=\protect\pi ^{2/3}$, corresponds to the situation
presented in Fig.~\protect\ref{fig:CP&EPS_SD}. The solid segment of the
yellow line below the open circle indicates the stable stationary SD
solitons with real wavefunctions. The dashed segment above the circle
represents the stable stationary solitons with the complex wavefunctions.}
\label{fig:Ncr123_SD}
\end{figure}

Figure~\ref{fig:Ncr123_SD} shows the three bifurcation values $N_{\text{cr}%
}^{(1,2,3)}$ as functions of the coupling parameter $g$ and distinct
dynamical states separated by them:

(i) Real-SD monostability (the red region below the solid curve): Only
real-wavefunction SD solitons exist stably and are the GSs in this region,
where the SD solitons with the complex stationary wavefunctions do not exist.

(ii) The oscillatory instability of the SD solitons with the real stationary
wavefunctions transforms them into breathers, such as the one displayed in
Fig.~\ref{fig:2UnStaSD}(d) (the blue region above the solid curve).
In this region, the stable SD solitons with the complex
stationary wavefunctions, which represent the GS, coexist with the
oscillatory SD solitons.

(iii) The splitting instability affects stationary SD solitons
with real wavefunctions, as exemplified in Fig.~\ref{fig:2UnStaSD}(h) (the 
brown area). In contrast, stationary SD solitons with complex wavefunctions, 
which serve as the GSs in this regime, remain stable throughout this parameter
region.

(iv) In the gray region above the dashed-dotted curve, solely the stable
stationary solitons with the complex wavefunctions exist as SD modes.

The analysis of the 2D SOC model with the MF cubic self-attraction
had revealed, apart from the two-component semi-vortex solitons, whose 1D
counterparts are the SD modes considered above, a family of stationary
solutions constitutes the mixed-mode (MM) solitons, characterized by the
combination of zero-vorticity and vortex components in both wavefunctions
\cite{Sakaguchi2014}. In the present 1D setting, two-component solutions
of the MM type can be generated by the following input:

\begin{eqnarray}
	\psi _{1}(t &=&0)=(A_{1}+A_{2}x)\exp (-x^{2}/l_{0}^{2}),  \label{MM1} \\
	\psi _{2}(t &=&0)=(A_{1}-A_{2}x)\exp (-x^{2}/l_{0}^{2}),  \label{MM2}
\end{eqnarray}%
with $A_{1}=1$, $A_{2}=0.5$ and $l_{0}=\sqrt{2}\approx \allowbreak 1.41$,
cf. Eqs.~(\ref{SD1}) and (\ref{SD2}). We numerically solved the simplified 
Eq.~(\ref{eq:GPEs_1}) in the LHY-only regime with input (\ref{MM1}), 
(\ref{MM2}) by means of AITEM. However, all stationary MM solutions admit 
strictly real wavefunctions up to a constant global phase. We have concluded 
that no genuinely complex MM states exist, any phase structure reducing to 
the trivial phase rotation. Consequently, while stable MM solitons exist, 
with properties similar to those of the SD states, they do not exhibit the 
symmetry-breaking bifurcation observed in the SD family.

\subsection{The LHY superfluid}

Many of the above results are produced for the value of the mean-field
interaction coefficient $g=\pi ^{2/3}$, which adequately represents the
generic case. Another interesting case corresponds to the limit of $%
g\rightarrow \infty $. In this case, applying the rescaling, $\psi
_{1,2}\equiv \sqrt{\pi }g^{-3/4}\Psi _{1,2}$, one transforms the underlying
equations (\ref{eq:GPEs_1}) into the simplified GPE system, which is the
model of the LHY superfluid, in which the nonlinearity represented solely by
the LHY\ terms, cf. Ref. \cite{Jorgensen2018}:

\begin{equation}
\begin{aligned}
i\partial _{t}\Psi _{1}=&-\frac{1}{2}\partial _{xx}\Psi _{1}+\partial
_{x}\Psi _{2} \\
& -(|\Psi _{1}|^{2}+|\Psi _{2}|^{2})^{1/2}\Psi _{1},
\end{aligned}\label{eq:GPEs_LHY1}
\end{equation}

\begin{equation}
\begin{aligned}
	i\partial _{t}\Psi _{2}=&-\frac{1}{2}\partial _{xx}\Psi _{2}-\partial
_{x}\Psi _{1} \\
& -(|\Psi _{1}|^{2}+|\Psi _{2}|^{2})^{1/2}\Psi _{2}.
\end{aligned}\label{eq:GPEs_LHY2}
\end{equation}

The AITEM technique was employed to numerically solve Eqs.~(\ref%
{eq:GPEs_LHY1}) and (\ref{eq:GPEs_LHY2}), yielding stable soliton solutions
with a real chemical potential $\mu $ and real-valued stationary
two-component wavefunctions $\Phi _{1,2}(x)$ of two types, SD and MM ones,
while no solitons with complex stationary wavefunctions were found. This means
that the substitution of $\Psi_{1,2}=\exp \left( -i\mu t\right) \Phi_{1,2}(x)$
in Eqs.~(\ref{eq:GPEs_LHY1}) and (\ref{eq:GPEs_LHY2}) leads to the system of
real equations

\begin{eqnarray}
\mu \Phi _{1} &=&-\frac{1}{2}\Phi _{1}^{\prime \prime }+\Phi _{2}^{\prime
}-(\Phi _{1}^{2}+\Phi _{2}^{2})^{1/2}\Phi _{1},  \label{Phi1} \\
\mu \Phi _{2} &=&-\frac{1}{2}\Phi _{2}^{\prime \prime }-\Phi _{1}^{\prime
}-(\Phi _{1}^{2}+\Phi _{2}^{2})^{1/2}\Phi _{2}.  \label{Phi2}
\end{eqnarray}

\begin{figure}[tbp]
\includegraphics[width=6.5cm]{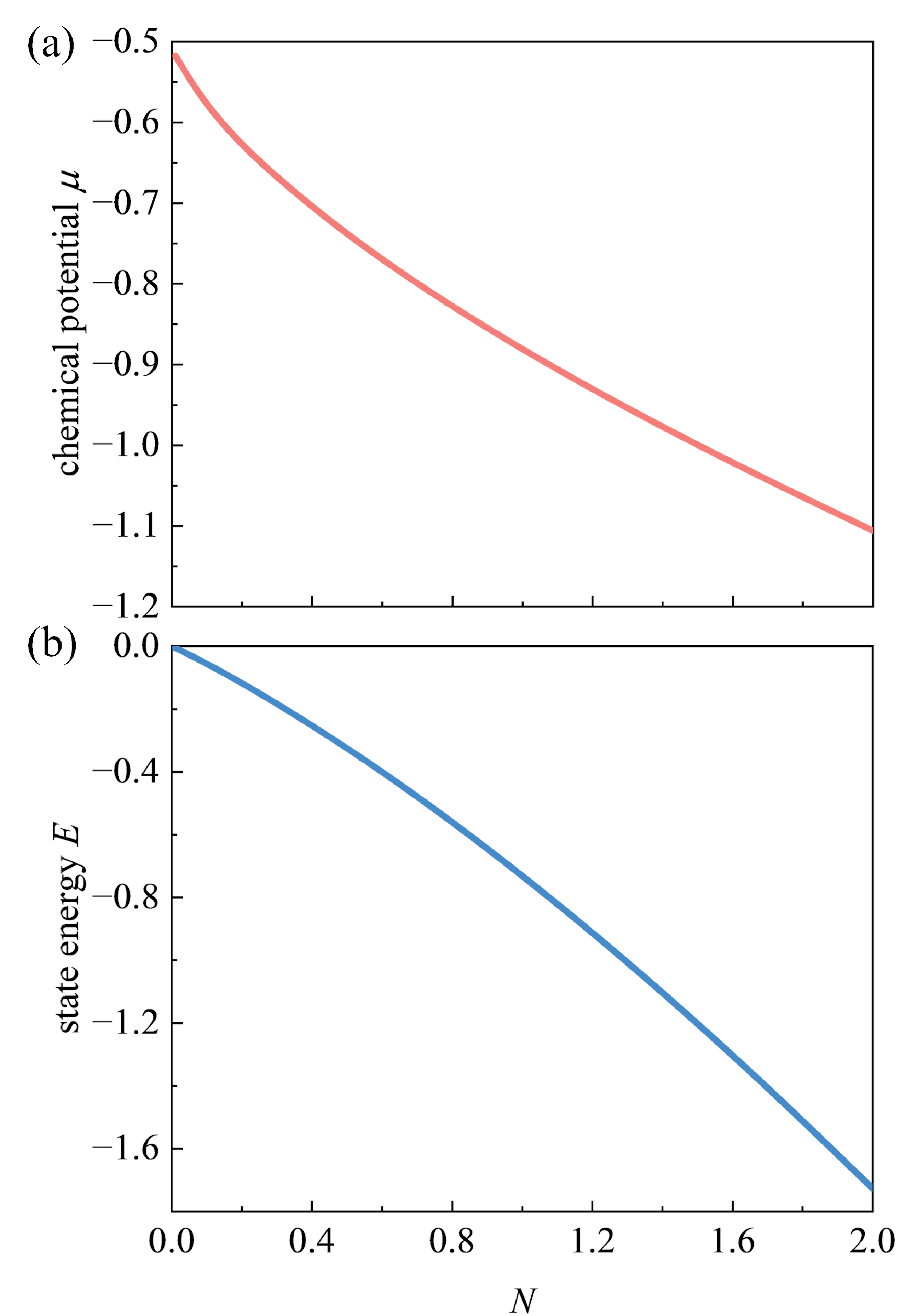}
\caption{(Color online) The degenerate (fully coinciding) chemical potential
and energy of the SD and MM solitons in the LHY superfluid, which
corresponds to $g\rightarrow \infty $. (a) Chemical potential $\protect\mu $
as a function of the soliton norm $N$. (b) Energy $E$ as a function of $N$.}
\label{fig:mu&E_LHY}
\end{figure}

The chemical potential $\mu $ and state energy $E$ of the SD and MM soliton
species in the LHY superfluid are plotted, as functions of the soliton norm $%
N$, in Fig.~\ref{fig:mu&E_LHY}. Numerical results reveal that both soliton
types, despite their starkly different spatial profiles, exhibit identical
values of $\mu $ and $E$ values at identical $N$, confirming their full
mutual degeneracy, underscoring the thermodynamic equivalence of the two
soliton families in the LHY regime.

\begin{figure*}[tbp]
\includegraphics[width=16.5cm]{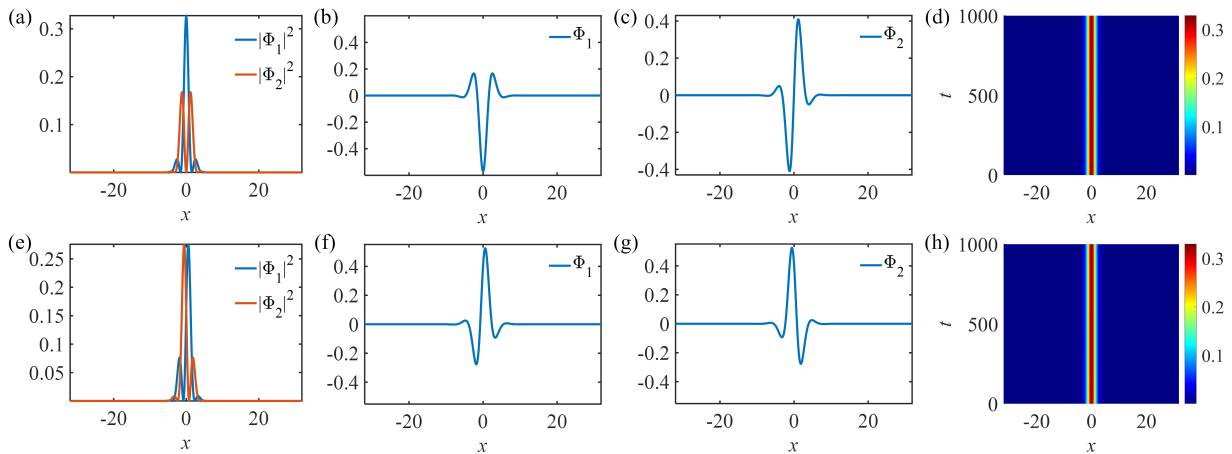}
\caption{(Color online) The coexistence between the two species of stable 
solitons with real stationary wavefunctions and norm $N=1$ in the 1D SOC 
LHY superfluid. Panels (a-d) represent the SD (semi-dipole) soliton: (a) 
the symmetric density profile; (b,c) the even and odd wavefunction components.
(d) Real-time evolution under a 1\% random perturbation 
applied to the initial stationary state, demonstrating that the SD soliton 
remains robust. Panels (e-h) represent the MM (mixed-mode) soliton: (e) 
the mirror-symmetric dual-lobe density structure; (f,g) real wavefunctions 
of both components, featuring the mutual mirror symmetry. (h) 
Real-time evolution under the same perturbation, confirming the dynamical 
stability of the MM soliton.} \label{fig:2Sta_SD&MM_LHY}
\end{figure*}

Figure~\ref{fig:2Sta_SD&MM_LHY} depicts examples of the stable
real-wavefunction SD and MM solitons in the LHY superfluid, both with norm $%
N=1$ but exhibiting fundamentally different shapes. The SD soliton displays
a symmetric dual-lobe density profile in panel (a), with the characteristic
structure of its components, \textit{viz}., spatially even $\Phi _{1}(x)$
and odd $\Phi _{2}(x)$, as seen in panels (b) and (c), respectively. 
Its robustness was verified by real-time simulations in panel 
(d) under a 1\% random perturbation applied to the initial stationary state, 
corroborating the stability of the SD soliton.

The stationary real wavefunctions of the MM soliton also 
underwent the same stability test under a 1\% random perturbation, maintains 
full mirror symmetry between its components in Figs.~\ref{fig:2Sta_SD&MM_LHY}%
(f,g), while its density profile features the identical dual-lobe structure 
in both components, as seen in panel (e). The evolution in 
panel (h) confirms that the MM soliton remains stable under such perturbations.

The coexistence of these\ stable solitons underscores their dichotomy in 
the LHY regime: while the spatial configurations and symmetries of the SD 
and MM families are widely different, both are completely stable, sharing 
the same value of the energy for all values of the norm. This duality highlights 
the structural diversity admitted by the LHY superfluids, even within the
constraint of the real-values stationary wavefunctions.

\begin{figure}[tbp]
\includegraphics[width=6.5cm]{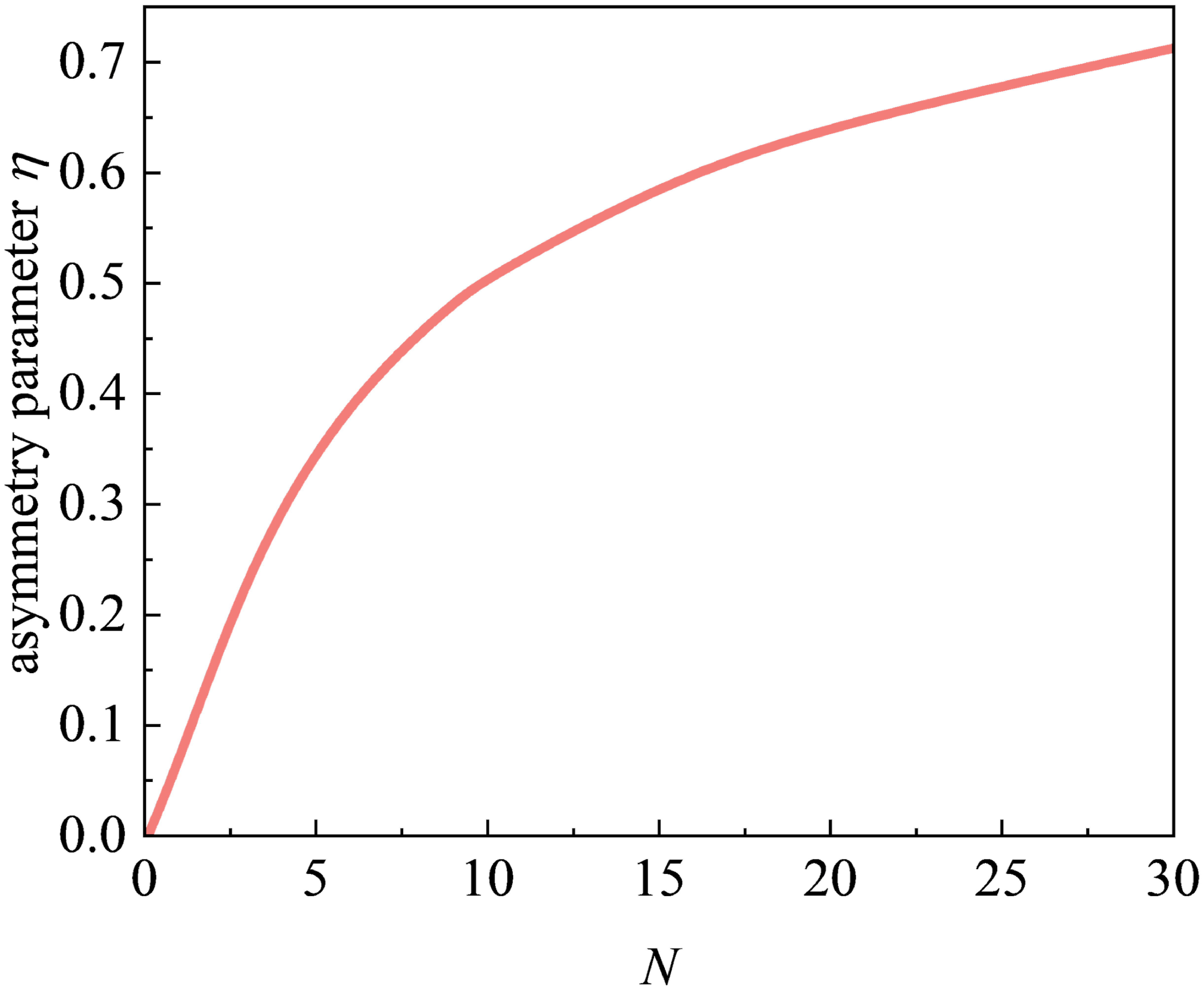}
\caption{(Color online) The inter-component norm-asymmetry parameter (\protect
\ref{eta}) of the the SD-type solitons, in the LHY-superfluid regime, vs. the
total norm $N$. The dependence exhibits a two-stage shape: the rapid initial
growth and saturation at larger $N$, highlighting the dominance of the
even-parity component for larger nonlinearity.} \label{fig:eta_LHY}
\end{figure}

In contrast to the MM solitons that maintain equal norms of both components,
the SD solitons exhibit asymmetry in this respect: as the total norm $N$
increases, the norm-asymmetry measure $\eta $, defined as per Eq.~(\ref{eta}%
), increases too, as shown in Fig.~\ref{fig:eta_LHY}. The dependence of $%
\eta (N)$ reveals the two-stage evolution with the growth of $N$: an initial
rapid growth of $\eta $ is followed by a saturation regime at larger $N$.
The latter feature can be easily explained by an asymptotic analysis of Eqs.~%
(\ref{Phi1}) and (\ref{Phi2}). Indeed, in the lowest approximation, the
solution for $N\rightarrow \infty $ degenerates into one with the vanishing
odd component, $\Phi _{2}\rightarrow 0$, and the even component represented
by the simple soliton solution of Eq.~(\ref{Phi1}) with $\Phi _{2}=0$ and a
simple expression for the norm:%
\begin{equation}
\Phi _{1}(x)=\frac{-3\mu }{2\cosh ^{2}\left( \sqrt{-\mu /2}x\right) },~N=3%
\sqrt{2}\left( -\mu \right) ^{3/2}.  \label{lowest}
\end{equation}%
The first correction to this solution is given by the odd mode determined 
by the simplified version of Eq.~(\ref{Phi2}), linearized with respect to 
$\Phi_{2}$:%
\begin{equation}
\frac{1}{2}\Phi _{2}^{\prime \prime }+\left[ \mu +\Phi _{1}(x)\right] \Phi
_{2}=-\Phi _{1}^{\prime }(x).  \label{linear}
\end{equation}%
A simple estimate demonstrates that Eq.~(\ref{linear}) yields a solution
with amplitude $\left( \Phi _{2}\right) _{\max }\sim \sqrt{-\mu }\sim N^{1/3}
$, hence the norm of this component is estimated as $N_{2}\sim N^{1/3}$. The
substitution of this in Eq.~(\ref{eta}) yields $1-\eta \equiv 2N_{2}/N\sim
N^{-2/3}$, which provides the explanation of the shape of the curve in Fig.~%
\ref{fig:eta_LHY} for large values of $N$.

\section{Conclusion}

In this work, we have systematically explored the existence, stability, and
dynamics of soliton solutions in the 1D SOC (spin-orbit-coupled) binary BEC
system incorporating the LHY (Lee-Huang-Yang) quantum-fluctuation corrections.
Our analysis focuses primarily on semi-dipole (SD) solitons, exhibiting
distinct structural transitions and stability regimes governed by nonlinear
interactions and SOC effects. Thus, we have demonstrated that the SD
solitons undergo the supercritical bifurcation (the phase transition of the
second kind) from purely real to complex wavefunctions, following the
increase of the norm. A corresponding phase diagram of the system reveals
the intricate stability landscape, including transitions of the SD solitons
to oscillatory and splitting instabilities.

In the LHY-dominated regime, corresponding to the limit of diverging
contact-interaction strength, $g\rightarrow \infty $, we have identified a
special class of SD and MM soliton solutions characterized by strictly real
wavefunctions. Notably, the SD and MM solitons coexist as stable modes with
identical chemical potentials and total energies at fixed values of the
total norm. This degeneracy highlights the thermodynamic equivalence and
structural diversity inherent in the LHY superfluid, being a unique aspect
of the quantum matter dominated by the LHY nonlinearity (quantum
fluctuations).

Thus, our findings underscore the critical role of quantum fluctuations,
encapsulated by LHY correction, in stabilizing self-bound modes in ultracold
quantum gases. The emergence of stable complex-valued solitons in this
context exhibits a significant departure from the conventional scalar
soliton phenomenology, highlighting the interplay between the spinor
effects, SOC, and LHY corrections.

The present work suggests directions for further theoretical and
experimental investigations. These include soliton collisions, the impact of
external trapping potentials, and extensions to higher-dimensional settings.
The demonstrated bistability and symmetry-breaking phenomena suggest
possible applications to quantum data processing and coherent matter-wave
control, where the use of controllable multistable states is highly
relevant.

%\appendix
%\section{}

\begin{acknowledgments}
The work of G.H.C. is supported by the Guangdong Basic and Applied Basic
Research Foundation (Grant No. 2024A1515010710). The work of H.C.W. is
supported by Dongguan Science and Technology of Social Development Program
(Grant No. 20231800940532), Songshan Lake Sci-Tech Commissioner Program
(Grant No. 20234373–01KCJ-G). The work of H.M.D. is supported by the 
Project of the Natural Science Foundation of Hunan Province, China (Grant 
No. 2024JJ5364) and the Hunan Provincial Education Office (Grant No. 23A0593). 
The work of B.A.M. is supported, in part, by the Israel Science Foundation 
(Grant No. 1695/22).
\end{acknowledgments}

% \bibliographystyle{plain}
% \bibliography{basename of .bib file}

\begin{thebibliography}{99}
\bibitem{Galitski2013} V. Galitski and I. B. Spielman, ``Spin-orbit coupling
in quantum gases,'' \textit{Nature}~\textbf{494}, 49--54 (2013).
DOI:10.1038/nature11841.

\bibitem{Zhang2016} Y. Zhang, M. E. Mossman, T. Busch, P. Engels, and C.
Zhang, ``Properties of spin-orbit-coupled Bose-Einstein condensates,''
\textit{Front. Phys.}~\textbf{11}, 118103 (2016).
DOI:10.1007/s11467-016-0560-y.

\bibitem{Zhai2015} H. Zhai, ``Degenerate quantum gases with spin-orbit
coupling: A review,'' \textit{Rep. Prog. Phys.} \textbf{78}, 026001 (2015).
DOI:10.1088/0034-4885/78/2/026001.

\bibitem{Lin2011} Y. J. Lin, K. Jim\'{e}nez-Garc\'{\i}a, and I. B. Spielman,
``Spin-orbit-coupled Bose-Einstein condensates,'' \textit{Nature}~\textbf{471}, 
83--86 (2011). DOI:10.1038/nature09887.

\bibitem{Ji2014} S.-C. Ji, J.-Y. Zhang, L. Zhang, Z.-D. Du, W. Zheng, Y.-J.
Deng, H. Zhai, S. Chen, and J.-W. Pan, ``Experimental determination of the
finite-temperature phase diagram of a spin-orbit coupled Bose gas,'' 
\textit{Nat. Phys.}~\textbf{10}, 314--320 (2014). DOI:10.1038/nphys2905.

\bibitem{Goldman2014} N. Goldman, G. Juzeli\={u}nas, P. \"{O}hberg, and I.
B. Spielman, ``Light-induced gauge fields for ultracold atoms,'' 
\textit{Rep. Prog. Phys.}~\textbf{77}, 126401 (2014).
DOI:10.1088/0034-4885/77/12/126401.

\bibitem{Dalibard2011} J. Dalibard, F. Gerbier, G. Juzeli\={u}nas, and P.
\"{O}hberg, ``Colloquium: Artificial gauge potentials for neutral atoms,''
\textit{Rev. Mod. Phys.}~\textbf{83}, 1523--1543 (2011).
DOI:10.1103/RevModPhys.83.1523.

\bibitem{Sakaguchi2014} H. Sakaguchi, B. Li, B. A. Malomed, ``Creation of
two-dimensional composite solitons in spin-orbit-coupled self-attractive
Bose--Einstein condensates in free space,'' \textit{Phys. Rev. E}~\textbf{89}, 
032920 (2014). DOI:10.1103/physreve.89.032920.

\bibitem{Malomed2016} B. A. Malomed, ``Multidimensional solitons:
Well-established results and novel findings,'' \textit{Eur. Phys. J. Spec.
Top.}~\textbf{225}, 2507--2532 (2016).
DOI:10.1140/epjst/e2016-60025-y.

\bibitem{Li2018} Y. Li, Z. Chen, Z. Luo, C. Huang, H. Tan, W. Pang, and B.
A. Malomed, ``Two-dimensional vortex quantum droplets,'' 
\textit{Phys. Rev. A}~\textbf{98}, 063602 (2018). 
DOI:10.1103/PhysRevA.98.063602.

\bibitem{Sakaguchi2018} H. Sakaguchi and B. A. Malomed, ``One- and
two-dimensional gap solitons in spin-orbit-coupled systems with Zeeman
splitting,'' \textit{Phys. Rev. A}~\textbf{97}, 013607 (2018).
DOI:10.1103/PhysRevA.97.013607.

\bibitem{Tononi2019} A. Tononi and L. Salasnich, ``Bose--Einstein
condensation on the surface of a sphere,'' \textit{Phys. Rev. Lett.}~\textbf{123}, 
160403 (2019). DOI:10.1103/PhysRevLett.123.160403.

\bibitem{Kevrekidis2015} P. G. Kevrekidis and D. J. Frantzeskakis,
``Solitons in coupled nonlinear Schr\"{o}dinger models: A
survey of recent developments,'' \textit{Rev. Phys.}~\textbf{1}, 
140-153 (2016). DOI:10.1016/j.revip.2016.07.002.

\bibitem{Khaykovich2002} L. Khaykovich, F. Schreck, G. Ferrari, T. Bourdel,
J. Cubizolles, L. D. Carr, Y. Castin, and C. Salomon, ``Formation of a
matter-wave bright soliton,'' \textit{Science}~\textbf{296}, 1290--1293 (2002). 
DOI:10.1126/science.1071021.

\bibitem{Strecker2002} K. E. Strecker, G. B. Partridge, A. G. Truscott, and
R. G. Hulet, ``Formation and propagation of matter-wave soliton trains,''
\textit{Nature}~\textbf{417}, 150--153 (2002).
DOI:10.1038/nature747.

\bibitem{Nguyen2014} J. H. V. Nguyen, P. Dyke, D. Luo, B. A. Malomed, and R.
G. Hulet, ``Collisions of matter-wave solitons,'' \textit{Nat. Phys.}~\textbf{10}, 
918--922 (2014). DOI:10.1038/nphys3135.

\bibitem{Kartashov2019} Y. V. Kartashov and D. A. Zezyulin, ``Stable
multiring and rotating solitons in two-dimensional spin-orbit-coupled
Bose--Einstein condensates with a radially periodic potential,'' 
\textit{Phys. Rev. Lett.}~\textbf{122}, 123201 (2019).
DOI:10.1103/PhysRevLett.122.123201.

\bibitem{Malomed2005} B. A. Malomed, D. Mihalache, F. Wise, and L. Torner,
``Spatiotemporal optical solitons,'' \textit{J. Opt. B: Quantum Semiclass.
Opt.}~\textbf{7}, R53--R72 (2005). DOI:10.1088/1464-4266/7/5/R02.

\bibitem{Sakaguchi2014a} H. Sakaguchi and B. A. Malomed, ``Discrete and
continuum composite solitons in Bose-Einstein condensates with the Rashba
spin-orbit coupling in one and two dimensions,'' 
\textit{Phys. Rev. E}~\textbf{90}, 062922 (2014). DOI:10.1103/PhysRevE.90.062922.

\bibitem{Kartashov2013} Y. V. Kartashov, V. V. Konotop, and F. K. Abdullaev,
``Gap solitons in a spin-orbit-coupled Bose-Einstein condensate,'' 
\textit{Phys. Rev. Lett.}~\textbf{111}, 060402 (2013).
DOI:10.1103/PhysRevLett.111.060402.

\bibitem{Kartashov2014} Y. V. Kartashov, V. V. Konotop, D. A. Zezyulin,
``Bose-Einstein condensates with localized spin-orbit coupling: Soliton
complexes and spinor dynamics,'' \textit{Phys. Rev. A}~\textbf{90}, 063621
(2014). DOI:10.1103/PhysRevA.90.063621.

\bibitem{Lobanov2014} V. E. Lobanov, Y. V. Kartashov, and V. V. Konotop,
``Fundamental, multipole, and half-vortex gap solitons in spin-orbit coupled
Bose-Einstein condensates,'' \textit{Phys. Rev. Lett.}~\textbf{112}, 180403
(2014). DOI:10.1103/physrevlett.112.180403.

\bibitem{Gautam2017} S. Gautam and S. K. Adhikari, ``Vortex-bright solitons
in a spin-orbit-coupled spin-1 condensate,'' \textit{Phys. Rev. A}~\textbf{95}, 
013608 (2017). DOI:10.1103/PhysRevA.95.013608.

\bibitem{Xu2013} Y. Xu, Y. Zhang, and B. Wu, ``Bright solitons in
spin-orbit-coupled Bose-Einstein condensates,'' \textit{Phys. Rev. A}~\textbf{87}, 
013614 (2013). DOI:10.1103/PhysRevA.87.013614.

\bibitem{Achilleos2013} V. Achilleos, D. J. Frantzeskakis, P. G. Kevrekidis,
and D. E. Pelinovsky, ``Matter-wave solitons in spin-orbit-coupled
Bose--Einstein condensates,'' \textit{Phys. Rev. Lett.}~\textbf{110}, 264101
(2013). DOI:10.1103/PhysRevLett.110.264101.

\bibitem{Sakaguchi2014b} H. Sakaguchi and B. A. Malomed, ``Matter-wave
solitons in nonlinear optical lattices,'' \textit{Phys. Rev. E}~\textbf{72},
046610 (2005). DOI:10.1103/physreve.72.046610.

\bibitem{Merkl2010} M. Merkl, A. Jacob, F. E. Zimmer, P. \"{O}hberg, and L.
Santos, ``Chiral confinement in quasirelativistic Bose-Einstein
condensates,'' \textit{Phys. Rev. Lett.}~\textbf{104}, 073603 (2010).
DOI:10.1103/PhysRevLett.104.073603.

\bibitem{Petrov2016} D. S. Petrov and G. E. Astrakharchik, ``Ultradilute 
low-dimensional liquids,'' \textit{Phys. Rev. Lett.}~\textbf{117}, 100401 
(2016). DOI:10.1103/PhysRevLett.117.100401.

\bibitem{Petrov2015} D. S. Petrov, ``Quantum mechanical stabilization of 
a collapsing Bose-Bose mixture,'' \textit{Phys. Rev. Lett.}~\textbf{115}, 
155302 (2015). DOI:10.1103/PhysRevLett.115.155302.

\bibitem{Astrakharchik2018} G. E. Astrakharchik and B. A. Malomed,
``Dynamics of one-dimensional quantum droplets,'' 
\textit{Phys. Rev. A}~\textbf{98}, 013631 (2018). DOI:10.1103/PhysRevA.98.013631.

\bibitem{Otajonov2019} S. R. Otajonov, E. N. Tsoy, and F. Kh. Abdullaev,
``Modulational instability and quantum droplets in a two-dimensional 
Bose--Einstein condensate,'' \textit{Phys. Rev. A}~\textbf{106}, 
033309 (2022). DOI:10.1103/physreva.106.033309.

\bibitem{Tylutki} M. Tylutki, G. E. Astrakharchik, B. A. Malomed, and D. S.
Petrov, ``Collective excitations of a one-dimensional quantum droplet,'' 
\textit{Phys. Rev. A}~\textbf{101}, 051601(R) (2020). 
DOI: 10.1103/physreva.101.051601.

\bibitem{Cabrera2018} C. R. Cabrera, L. Tanzi, J. Sanz, B. Naylor, P.
Thomas, P. Cheiney, and L. Tarruell, ``Quantum liquid droplets in a mixture
of Bose-Einstein condensates,'' \textit{Science}~\textbf{359}, 301--304 (2018). 
DOI:10.1126/science.aao5686.

\bibitem{Semeghini2018} G. Semeghini, G. Ferioli, L. Masi, C. Mazzinghi, L.
Wolswijk, F. Minardi, M. Modugno, G. Modugno, M. Inguscio, and M. Fattori,
``Self-bound quantum droplets of atomic mixtures in free space,'' 
\textit{Phys. Rev. Lett.}~\textbf{120}, 235301 (2018).
DOI:10.1103/PhysRevLett.120.235301.

\bibitem{DErrico2019} C. D'Errico, A. Burchianti, M. Prevedelli, L.
Salasnich, F. Ancilotto, M. Modugno, F. Minardi, and C. Fort,
``Observation of quantum droplets in a heteronuclear bosonic
mixture,'' \textit{Phys. Rev. Res.}~\textbf{1}, 033155 (2019).
DOI:10.1103/PhysRevResearch.1.033155.

\bibitem{Ferrier2016} I. Ferrier-Barbut, H. Kadau, M. Schmitt, M. Wenzel,
and T. Pfau, ``Observation of quantum droplets in a strongly dipolar Bose
gas,'' \textit{Phys. Rev. Lett.}~\textbf{116}, 215301 (2016).
DOI:10.1103/PhysRevLett.116.215301.

\bibitem{Schmitt2016} M. Schmitt, M. Wenzel, F. B\"{o}ttcher, I.
Ferrier-Barbut, and T. Pfau, ``Self-bound droplets of a
dilute magnetic quantum liquid,'' \textit{Nature}~\textbf{539}, 
259--262 (2016). DOI:10.1038/nature20126.

\bibitem{Baillie2016} D. Baillie, R. M. Wilson, R. N. Bisset, and P. B.
Blakie, ``Collective excitations of self-bound droplets of a dipolar quantum
fluid,'' \textit{Phys. Rev. A} \textbf{94}, 021602 (2016).
DOI:10.1103/PhysRevA.94.021602.

\bibitem{Bottcher2021} F. B\"{o}ttcher, J.-N. Schmidt, J. Hertkorn, K. S. H.
Ng, S. D. Graham, M. Guo, T. Langen, and T. Pfau, ``New states of matter
with fine-tuned interactions: quantum droplets and dipolar supersolids,''
\textit{Rep. Prog. Phys.}~\textbf{84}, 012403 (2021).
DOI:10.1088/1361-6633/abc9ab.

\bibitem{Cheiney2018} P. Cheiney, C. R. Cabrera, J. Sanz, B. Naylor, L.
Tanzi, and L. Tarruell, ``Bright soliton to quantum droplet transition in a
mixture of Bose--Einstein condensates,'' \textit{Phys. Rev. Lett.}~\textbf{120}, 
135301 (2018). DOI:10.1103/PhysRevA.105.053616.

\bibitem{Cui2021} X. Cui and Y. Ma, ``Droplet under confinement: Competition
and coexistence with a soliton bound state,'' 
\textit{Phys. Rev. Research}~\textbf{3}, L012027 (2021). 
DOI:10.1103/PhysRevResearch.3.L012027.

\bibitem{Cappellaro2018} A. Cappellaro, T. Macr\'{\i}, and L. Salasnich,
`Collective modes across the soliton-droplet crossover in binary Bose
mixtures,'' \textit{Phys. Rev. A}~\textbf{97}, 053623
(2018). DOI:10.1103/PhysRevA.97.053623.

\bibitem{Gangwar2022} S. Gangwar, R. Ravisankar, P. Muruganandam, and P. K.
Mishra, ``Dynamics of quantum solitons in Lee-Huang-Yang spin-orbit-coupled
Bose-Einstein condensates,'' \textit{Phys. Rev. A}~\textbf{106}, 063315
(2022). DOI:10.1103/PhysRevA.106.063315.

\bibitem{Lee1957} T. D. Lee, K. Huang and C. N. Yang, ``Eigenvalues and
eigenfunctions of a Bose system of hard spheres and its low-temperature
properties,'' \textit{Phys. Rev.}~\textbf{106}, 1135 (1957).
DOI:10.1103/PhysRev.106.1135.

\bibitem{Jorgensen2018} N. B. J\o rgensen, G. M. Bruun, J. J. Arlt,
``Dilute fluid governed by quantum fluctuations,'' \textit{Phys. Rev. Lett.}
~\textbf{121}, 173403 (2018). DOI:10.1103/PhysRevLett.121.173403.

\bibitem{VakhKol} N. G. Vakhitov and A. A. Kolokolov, ``Stationary solutions 
of the wave equation in a medium with nonlinearity saturation,'' 
\textit{Radiophys. Quantum Electron.}~\textbf{16}, 783--789 (1973). 
DOI:10.1007/BF01031343.

\bibitem{Kuznetsov2011} E. A. Kuznetsov and F. Dias, ``Bifurcations of
solitons and their stability,'' \textit{Phys. Rep.}~\textbf{507}, 43--105
(2011). DOI:10.1016/j.physrep.2011.06.002.

\bibitem{Iooss1980} G. Iooss and D. D. Joseph, ``Elementary stability
bifurcation theory,'' Springer, New York, 1980.
\end{thebibliography}

\end{document}